\documentclass[aps,pra,reprint,amsmath,superscriptaddress,longbibliography]{revtex4-1}
\usepackage{natbib}

\usepackage[]{longtable}
\usepackage{graphicx}
\usepackage{todonotes}
\usepackage{braket}
\usepackage{amsmath}
\usepackage{here}
\usepackage{wasysym}

\begin{document}


\title{Plasma Propulsion of a Metallic Micro-droplet and its Deformation upon Laser Impact}



\author{Dmitry Kurilovich}
\affiliation{Advanced Research Center for Nanolithography (ARCNL), Science Park 110, 1098 XG Amsterdam, The Netherlands}
\affiliation{Department of Physics and Astronomy, and LaserLaB, Vrije Universiteit, De Boelelaan 1081, 1081 HV Amsterdam, The Netherlands}
\author{Alexander L. Klein}
\affiliation{Physics of Fluids Group, Faculty of Science and Technology, MESA+ Institute, University of Twente, P.O. Box 217, 7500 AE Enschede, The Netherlands}
\author{Francesco Torretti}
\affiliation{Advanced Research Center for Nanolithography (ARCNL), Science Park 110, 1098 XG Amsterdam, The Netherlands}
\affiliation{Department of Physics and Astronomy, and LaserLaB, Vrije Universiteit, De Boelelaan 1081, 1081 HV Amsterdam, The Netherlands}
\author{Adam Lassise}
\affiliation{ASML Netherlands B.V., De Run 6501, 5504 DR Veldhoven, The Netherlands}
\author{\mbox{Ronnie Hoekstra}}
\affiliation{Advanced Research Center for Nanolithography (ARCNL), Science Park 110, 1098 XG Amsterdam, The Netherlands}
\affiliation{Zernike Institute for Advanced Materials, University of Groningen, Nijenborgh 4, 9747 AG Groningen, The Netherlands}
\author{Wim Ubachs}
\affiliation{Advanced Research Center for Nanolithography (ARCNL), Science Park 110, 1098 XG Amsterdam, The Netherlands}
\affiliation{Department of Physics and Astronomy, and LaserLaB, Vrije Universiteit, De Boelelaan 1081, 1081 HV Amsterdam, The Netherlands}
\author{Hanneke Gelderblom}
\affiliation{Physics of Fluids Group, Faculty of Science and Technology, MESA+ Institute, University of Twente, P.O. Box 217, 7500 AE Enschede, The Netherlands}
\author{Oscar O. Versolato}
\affiliation{Advanced Research Center for Nanolithography (ARCNL), Science Park 110, 1098 XG Amsterdam, The Netherlands}
\email{o.versolato@arcnl.nl}


\date{\today}

\begin{abstract}
The propulsion of a liquid indium-tin micro-droplet by nanosecond-pulse laser impact is experimentally investigated. We capture the physics of the droplet propulsion in a scaling law that accurately describes the plasma-imparted momentum transfer, enabling the optimization of the laser-droplet coupling. The subsequent deformation of the droplet is described by an analytical model that accounts for the droplet's propulsion velocity and the liquid properties. Comparing our findings to those from vaporization-accelerated mm-sized water droplets, we demonstrate that the hydrodynamic response of laser-impacted droplets is scalable and independent of the propulsion mechanism. 
\end{abstract}


\maketitle

\section{Introduction}
Micro-droplets of tin can be used to create extreme ultra-violet light (EUV), \emph{e.g.} in next-generation nano-lithography machines \cite{Benschop2008,Banine2011}, where the droplets serve as mass-limited targets \cite{Fujioka2008,OSullivan2015} for a laser produced plasma (LPP). 
In such machines these droplets, several 10\,$\mu$m in diameter, are targeted by a focused nanosecond-pulse laser at intensities that lead to optical breakdown and the creation of a high-density and high-temperature plasma \cite{Banine2011,Tomita2015,OSullivan2015}. Line emission from electron-impact-excited highly charged tin ions in the plasma provides the EUV light \cite{Banine2011,OSullivan2015}.
For certain applications, it is beneficial to apply a dual-pulse sequence in which a laser ``prepulse'' is used to carefully shape the droplet into a thin sheet that is considered advantageous for EUV production with the subsequent, much more energetic, ``main pulse'' \cite{Fujioka2008,Mizoguchi2010,Banine2011}. 
Maximizing the conversion efficiency of such EUV sources, while minimizing the amount of fast ionic and neutral debris that limit EUV optics lifetime (see \cite{Mizoguchi2010,Coons2010,Banine2011} and references therein) requires a careful control over the droplet propulsion and its shape. This in turn requires a profound understanding of the coupling of the laser to the droplet as well as the droplet's fluid dynamic response to such a laser pulse impact.
While the propulsion and deformation of water droplets due to laser-pulse impact has been studied in detail \cite{Klein2015}, the interaction between liquid metal droplets and a pulsed laser has remained unexplored.\\ 

Here, we present an analysis of the response of liquid indium-tin droplets to laser-pulse impact in a setting very close to the one used in the generation of EUV light for the industrial application. We show that the droplet response to laser impact is governed by two physical processes that occur at completely different, and therewith separable, timescales.  First, the interaction of the nanosecond laser pulse with the metal generates a plasma \cite{Phipps1988,Fabbro1985} and induces propulsion of the droplet. Second, fluid dynamic effects govern the shape-evolution of the droplet, which takes place on a microsecond timescale. We reveal in detail the mechanism behind metal-droplet propulsion by laser impact and present a scaling law for the propulsion speed as a function of laser energy that captures the onset of plasma formation. This scaling law enables us to optimize the laser-droplet coupling for EUV generation purposes. Next, we discuss the similarities between metal- and water-droplet \cite{Klein2015} propulsion by laser impact. We demonstrate that even though the metal-droplet propulsion mechanism is completely different from that of water droplets, the fluid dynamics response is identical and well described by a universal analytical model.

\begin{figure}[b]
\includegraphics[width=8.6cm]{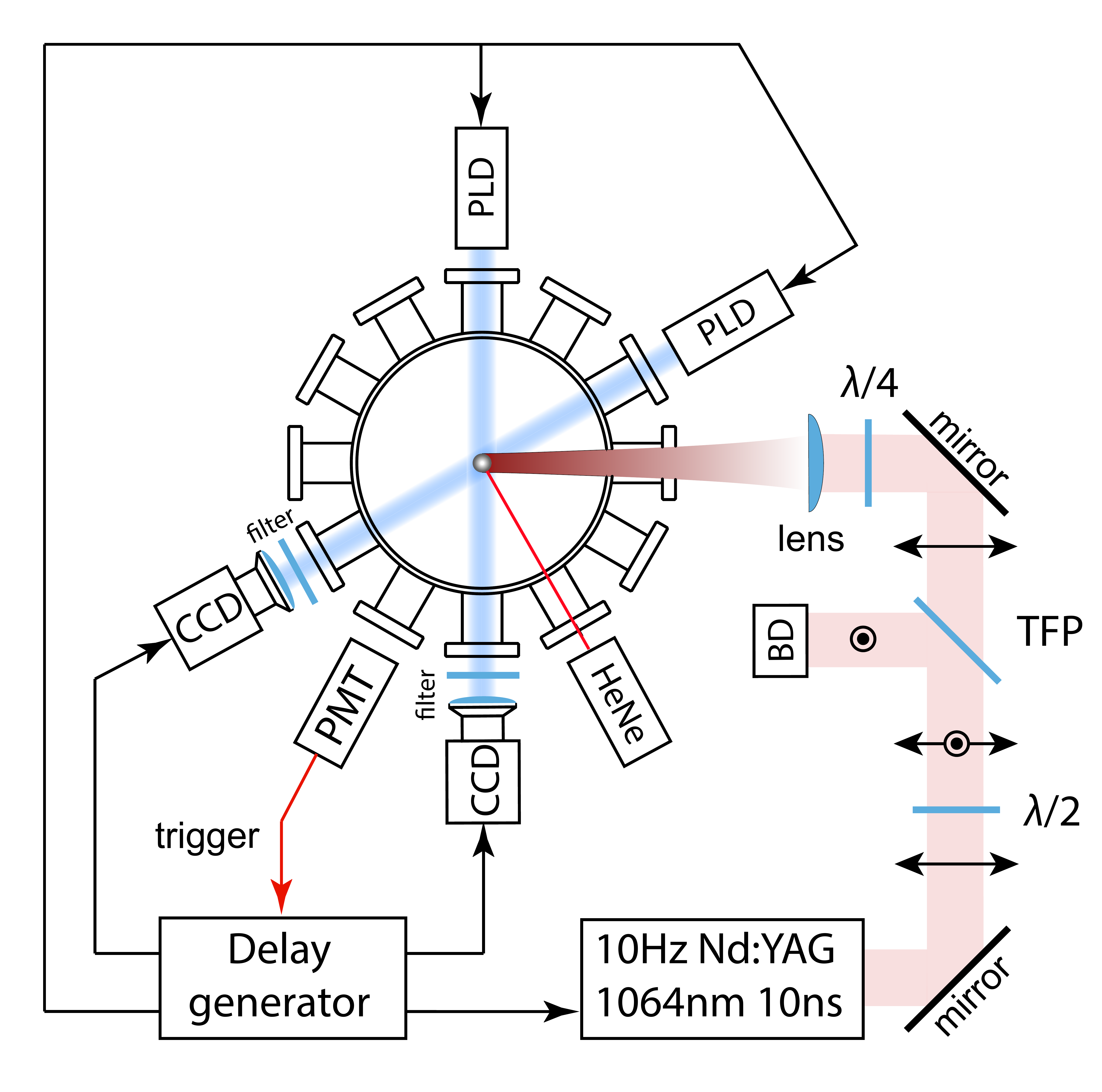}
\caption{Sketch of the experimental setup in top view. Depicted is a vacuum vessel, typically at $10^{-7}$\,mbar, with its many optical access ports. Droplets are dispensed from a nozzle (not shown) held at a temperature of 250$^{\circ}$\,C. A horizontal light sheet produced by a helium-neon (HeNe) laser enables the triggering of the experiment: the light scattered by the falling droplet is detected by a photomultiplier tube (PMT) that sends a signal to a delay generator. This delay generator in turn triggers an injection-seeded Nd:YAG laser that emits a pulse at $\lambda=1064\,$nm with a duration of $\tau_p \approx 10$\,ns full width at half maximum (FWHM). A 1000\,mm (500\,mm) focal-length N-BK7 lens produces a circular Gaussian focus of size $\sim$115\,$\mu$m (50\,$\mu$m) FWHM at the position of the droplet. The circle and arrow symbols along the optical axis indicate the direction of polarization of the laser beam. The energy of the laser pulse is controllable through the use of a half-wave plate ($\lambda/2$), a thin film polarizer (TFP), and a beam dump (BD). Before the focusing lens, a quarter-wave plate ($\lambda/4$) produces a circular polarization. The delay generator also triggers the pulsed laser diodes (PLD, at 850\,nm wavelength; $\sim$15\,ns pulse length) one of which is aligned orthogonally to the Nd:YAG laser light propagation direction. The other PLD is angled at 30 degrees to it. In both cases, the PLD light passes through a band pass filter before it falls onto the CCD chip of a camera through a long-distance microscope and produces a shadowgraph. The magnifications are 2.0(2)\,$\mu$m/pixel and 2.8(3)\,$\mu$m/pixel for the orthogonal and 30 degree shadowgraphs, respectively.
\label{fig:setup}}
\end{figure}%

\section{Experimental methods}
Figure\,\ref{fig:setup} shows a detailed description of the experimental set-up. We use a eutectic indium-tin alloy (50In-50Sn of 99.9\% purity with liquid density $\rho=6920$\,kg/m$^3$ at 250$^{\circ}$\,C \cite{Pstrus2013}), a substance almost equivalent to pure tin in terms of atomic mass, density, and surface tension, but with a conveniently low melting point. Droplets are dispensed from a droplet generator at $\sim$10\,kHz repetition rate, with final radii $R_0 \approx 25$\,$\mu$m, and are falling down at a speed of $\sim$12\,m/s in a vacuum vessel. No significant acceleration under gravity occurs on the time scale relevant for the experiment. The droplets relax to a spherical shape before they pass through a horizontal light sheet produced by a helium-neon laser. The light scattered by the droplets is used to trigger an injection-seeded Nd:YAG laser operating at 10\,Hz repetition rate that emits a pulse at its fundamental wavelength of $\lambda=1064\,$nm with a duration of $\tau_p \approx 10$\,ns full width at half maximum (FWHM), focused down to a circular Gaussian spot. We studied the propulsion for two different focusing conditions: $\sim$115\,$\mu$m and $\sim$50\,$\mu$m FWHM at the position of the droplet. In the first and main case of $\sim$115\,$\mu$m FHWM the width is larger than the droplet diameter in order to decrease the sensitivity to laser-to-droplet alignment and to provide a pressure profile required for obtaining a flat thin sheet \cite{Gelderblom:2015}. The finite geometrical overlap thus obtained precludes the full laser pulse energy from reaching the target. The energy of the laser pulse is varied in a manner that does not affect the transversal mode profile of the laser beam.
 
\begin{figure}[t]
\includegraphics[width=8.6cm]{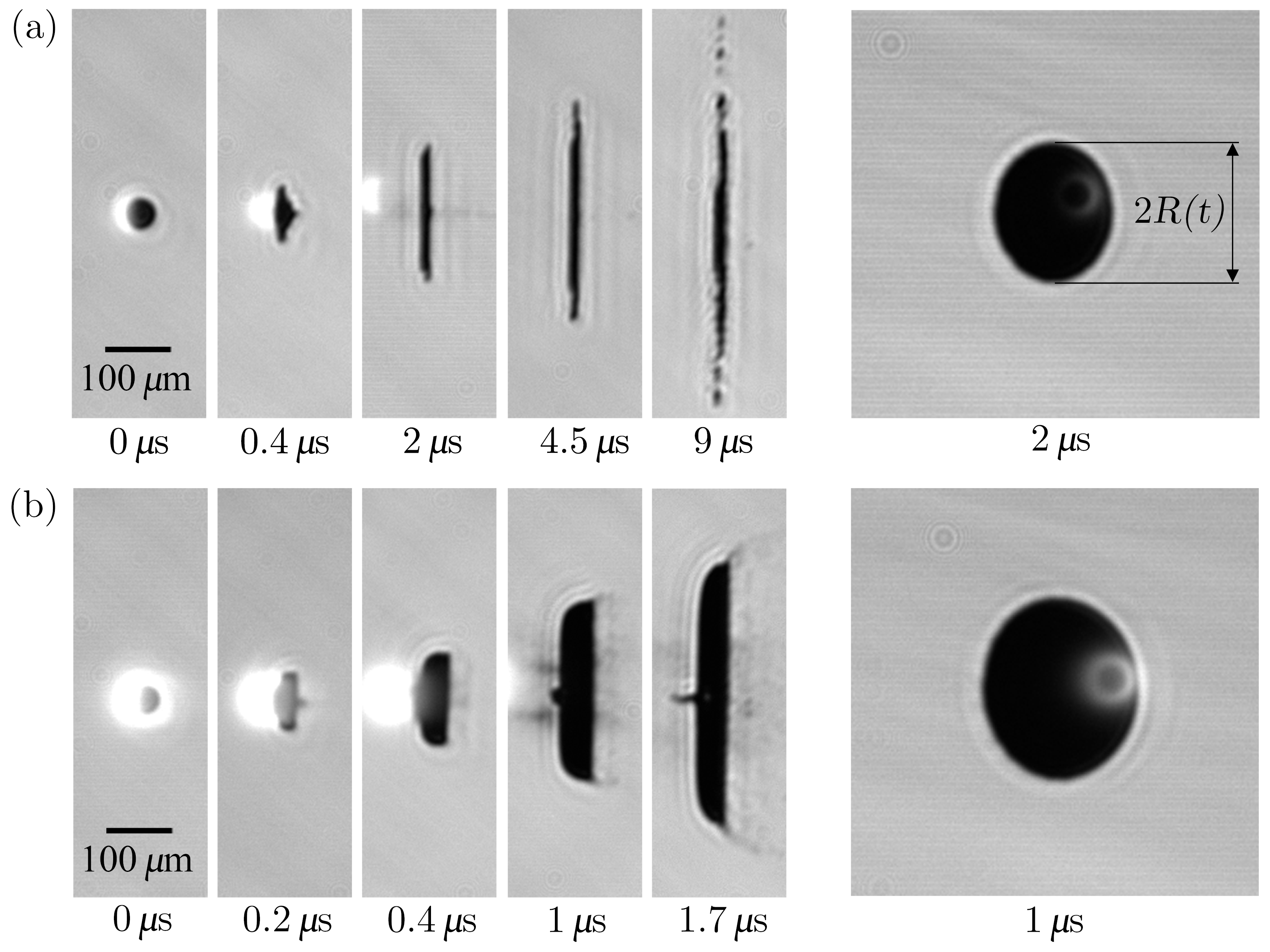}
\caption{Shadowgraphy images of In-Sn droplets in side and (tilted) front view (see Fig.\,\ref{fig:setup}). (a) Expansion of an In-Sn droplet as viewed from the side. Droplets are irradiated with a $\sim$115\,$\mu$m FWHM focused 10-ns Nd:YAG laser pulse impinging from the left with a total pulse energy of 7\,mJ. The stroboscopic droplet expansion sequence is constructed from shadowgraphs of different droplets at different time delays. A 30 degree view is provided on the right. (b) Same as (a) but for a 70\,mJ energy pulse, which gives rise to a faster expansion (note the different time scales). The white glow visible to the left of the expanding droplet is plasma light. Images have been vertically aligned to center on the expanding droplet.
\label{fig:droplets}}
\end{figure}

Imaging of the droplet is obtained by employing two pulsed laser diodes (PLDs). One of these is aligned orthogonally to the Nd:YAG laser light propagation direction to provide side-view images. The other PLD is angled at 30 degrees to it (see Fig.\,\ref{fig:setup}) for a (tilted) front view. In both cases, the PLD light passes through a band pass filter before it falls onto the CCD chip of a camera through a long-distance microscope. The shadowgraphic images thus obtained are used to track size, shape, and velocity of the droplet expansion in the direction along the Nd:YAG laser impact and perpendicular to the falling droplet. A stroboscopic time series of different droplets at various time delays (see Fig.\,\ref{fig:droplets}) with an arbitrary number of frames is constructed by triggering once per Nd:YAG shot, each time with an increasing delay.

\section{results}
The response of the In-Sn droplet to laser impact is shown in Fig.\,\ref{fig:droplets}. The laser pulse generates a plasma that expands away from the droplet surface. As a result, the droplet is accelerated to a speed $U\sim 0.5-350$\,m/s on a time $\tau_a$ given by the lifetime of the generated plasma, which is known to be smaller than a few $\tau_p$ \cite{Fabbro1990}. The subsequent deformation of the droplet occurs on the much longer inertial timescale  $\tau_i=R_0/U\approx  1\,\mu$s  (here, $U$ is taken from the case shown in Fig.\,\ref{fig:droplets}(a)). The deformation is eventually slowed down by surface tension on the capillary timescale $\tau_c=\sqrt{\rho R_0^3/\gamma}\approx 14\, \mu$s. The timescales relevant to this problem can thus be ordered\,\cite{Klein2015}
\begin{equation}
\tau_p < \tau_a \ll \tau_i < \tau_c,
\end{equation}
which illustrates the clear separation of timescales between plasma generation as a cause for the propulsion on the one side and the fluid-dynamic response on the other side. Below, we discuss first the droplet propulsion mechanism and second the droplet deformation.

\subsection{Droplet propulsion}
\begin{figure}[]
\includegraphics[width=8.6cm]{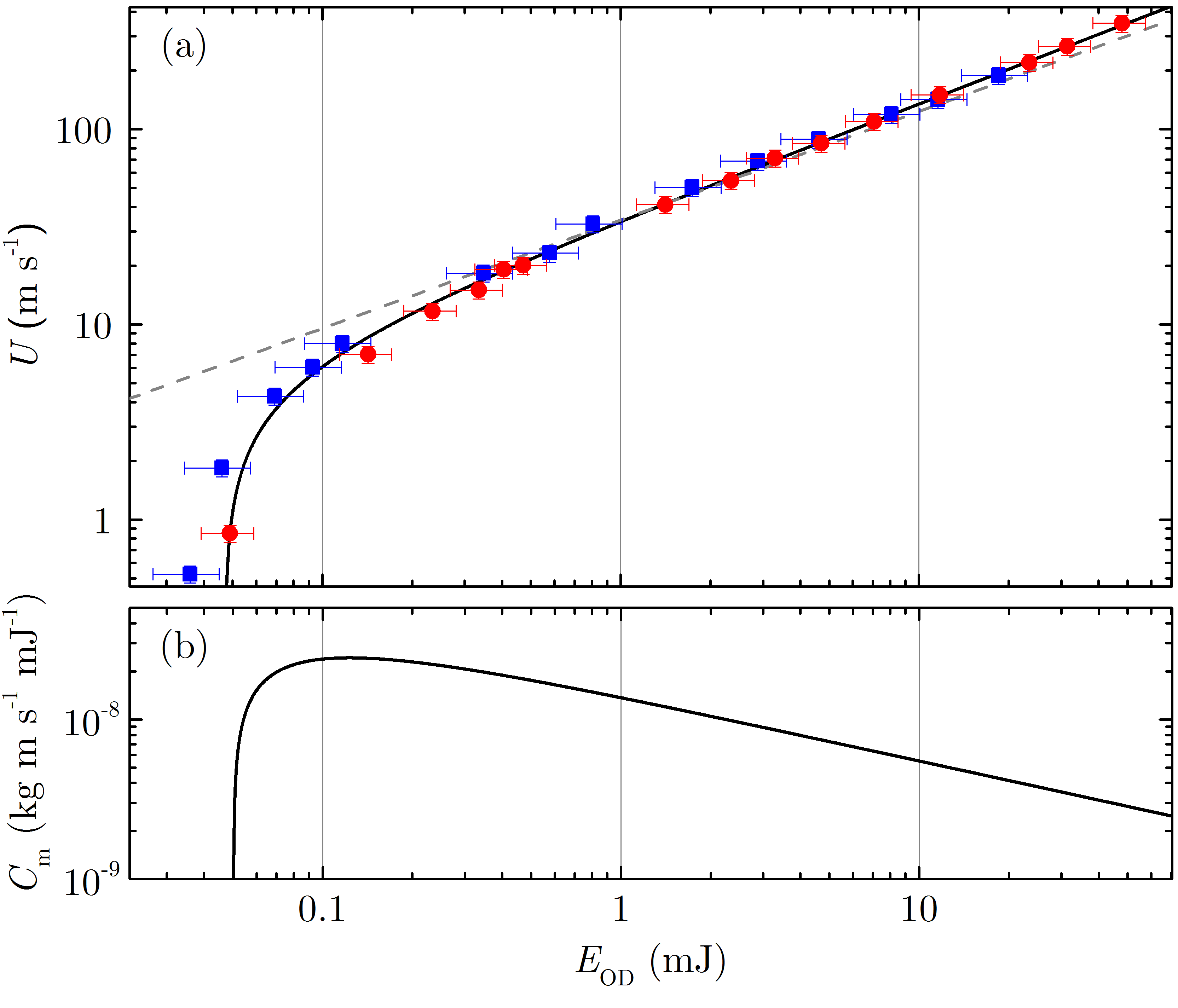}
\caption{(a) Propulsion velocity $U$ of In-Sn droplets as a function of total laser pulse energy impinging on the droplet $E_{\textrm{OD}}$ for two different focusing conditions. Blue squares represent the data obtained at a $\sim$115\,$\mu$m FWHM focus; red circles represent the $\sim$50\,$\mu$m case. Uncertainties ($\sim$10\%; 1$\sigma$) for the velocity are smaller than the symbol size; $E_{\textrm{OD}}$ is obtained with a $\sim$25\% (20\%) uncertainty in case of $\sim$115\,$\mu$m (50\,$\mu$m) FWHM focus. The gray dashed line represents a fit of Eq.\,(\ref{eq:powerlaw1}) to the concatenated data excluding pulse energies below 0.2\,mJ. The black solid line depicts a fit of Eq.\,(\ref{eq:powerlaw2}) to the concatenated data. (b) Momentum coupling coefficient $C_\textrm{m}$ obtained from the fitted curve shown as a black solid line in (a). 
\label{fig:powerlaw}}
\end{figure}

Using the shadowgraph images of the type shown in Fig.\,\ref{fig:droplets} we study the propulsion of the droplet as a function of the laser pulse energy, which is varied between 0.4 and 160\,mJ. Fig.\,\ref{fig:powerlaw} shows that over the nearly three decades of laser pulse energy studied in this work, the velocity of the droplet ranges from below 1\,m/s to above 300\,m/s and appears to follow a power-law type scaling with the laser pulse energy impinging on the droplet, $E_{\textrm{OD}}$. It is a fraction of the total pulse energy $E$ given by the geometrical overlap of laser focus and droplet. The similarity of the data obtained for the two different focusing conditions shows that $E_{\textrm{OD}}$ is indeed the relevant parameter describing the scaling of $U$ in the present study. Significant deviations from this simple parametrization are expected for foci much smaller than the droplet \cite{Gelderblom:2015}. To explain the observed scaling, we first discuss the mechanism through which the laser interacts with the metal.

The interaction of a high-intensity laser pulse with the droplet is governed by plasma dynamics \footnote{The typical impulse $p_l$ exerted on the droplet in the limit of fully absorbed light scales as $p_l \propto E/c$, with $E$ the energy of the absorbed light and $c$ the speed of light. Even at the highest pulse energies assuming a maximum fraction of absorbable light of $\sim$0.5, given by the geometrical overlap of laser and droplet in the case of the 50\,$\mu$m FWHM focus, this impulse would yield a typical droplet speed $U \approx 0.5$\,m/s which is negligible with respect to the observed speeds.}.
As soon as a plasma is generated, inverse bremsstrahlung absorption \cite{Mora1982} strongly decreases the initially high reflectivity of the metallic surface down to negligible values. This facilitates the further ablation of the target material. Analogous to the work on water droplets \cite{Klein2015}, we relate the propulsion speed $U$ of the micro-droplet (mass $M$) to the amount of ablated mass $m$ through momentum conservation 
\begin{equation}
M U = m v,
\label{momentumconservation}
\end{equation}
where $v$ is the velocity of the ejected mass along the axis of propulsion. The spatial distribution of this ejected mass peaks in the direction back towards the laser \cite{freeman2012laser}. In the present intensity regime, simulations predict $v$ to range from $5$-$15\times 10^3$\,m/s \cite{Masnavi2011} and to be a function of the laser intensity. Experimentally, $v \approx 8.5 \times 10^{3}$\,m/s at a laser intensity of $2\times 10^{11}$\,W/cm$^{2}$ \cite{Coons2010} which we take as a first estimate of the value for $v$ in the following. The ablated mass $m$ can be obtained employing a semi-empirical mass ablation law \cite{Harilal2011,Burdt2009},
\begin{equation}
m = A \times \pi R^2_0 \times \bigg[\bigg(\frac{I}{I_0}\bigg)^{5/9} \bigg(\frac{\lambda_0}{\lambda}\bigg)^{4/9} Z^{3/8}\bigg] \times \tau_p,
\label{rate}
\end{equation}
with laser pulse intensity $I$ and $I_0$=$10^{11}$\,W\,cm$^{-2}$, wavelength $\lambda$ and $\lambda_0$=$1\,\mu$m, and atomic number $Z$=49-50 (for indium-tin). The term in square brackets is based on an analytical treatment of the one-dimensional expansion of a plasma from a plane surface \cite{Dahmani1991}. The empirical constant $A$ was determined to be $3.0 \cdot 10^{3}$\,g\,cm$^{-2}$\,s$^{-1}$ for Sn for intensities ranging $10^{11}$-$10^{12}$\,W\,cm$^{-2}$\,\cite{Burdt2009}. Using this value for $A$ in Eq.\,(\ref{rate}) we obtain an ablated mass fraction of 0.1\% for the 7\,mJ example shown in Fig.\,\ref{fig:droplets}(a). However, the input for $A$ as well as for $v$ above was obtained experimentally at laser intensities one to two orders of magnitude higher than are used in the current investigations. Therefore, we will leave their product as a free fit parameter in the following.

The model (Eqs.\,(\ref{momentumconservation},\ref{rate})) predicts a power-law dependence of the propulsion velocity on the laser intensity, which can be translated into a dependence on the total laser energy impinging on the droplet $E_{\textrm{OD}}$ for direct comparison to our measurements by virtue of the relation $I \propto E_{\textrm{OD}}$, obtaining
\begin{equation}
U=B \cdot E_{\textrm{OD}}^{5/9}. \label{eq:powerlaw1}
\end{equation}
The result of a single fit of Eq.\,(\ref{eq:powerlaw1}) to both data sets excluding low pulse energies $E_{\textrm{OD}} < 0.2$\,mJ, yields excellent agreement of this power-law with the data as is shown in Fig.\,\ref{fig:powerlaw}. This agreement shows that $E_{\textrm{OD}}$ is the correct parameter in scaling the data for the two focusing conditions. We obtain a proportionality constant $B = 34(5)$\,m\,s$^{-1}$\,mJ$^{-5/9}$. Employing the empirical values from literature for $A$ and $v$ in Eqs.\,(\ref{momentumconservation},\ref{rate}) we obtain a prediction of $\sim$9\,m\,s$^{-1}$\,mJ$^{-5/9}$. A discrepancy was to be expected as the laser intensities in our experiments fall mostly outside the validated range \cite{Burdt2009} for Eq.\,(\ref{rate}) and for the estimate of $v$. Moreover, we disregarded the spherical geometry of the system. In light of these shortcomings, the agreement of the simple power law of Eq.\,(\ref{eq:powerlaw1}) with both data sets is striking.

Power-law scaling of the momentum imparted by an expanding plasma has been extensively studied in the context of plasma thrusters as well as nuclear fusion (e.g., see \cite{Phipps1988,Fabbro1985}). Empirically, from experiments on planar solid targets, it was found \cite{Phipps1988} that the ratio of the plasma pressure $p$ and laser intensity $I$ was excellently reproduced by the relation 
\begin{equation}
p/I \propto  (I \lambda \sqrt{\tau_p})^{n},\label{pressurelaw} 
\end{equation}
with exponent $n=-0.30(3)$ common to all studied materials \cite{Phipps1988,Fabbro1985}. This scaling law was found to be valid over a very broad range of target and laser parameters, including the intensities and wavelengths relevant for EUV sources. Employing the scaling relation $U \propto I^{5/9}$ from Eq.\,(\ref{eq:powerlaw1}) and linearizing $U \propto p \cdot \tau_p$, we obtain $n=-4/9 \approx -0.44$ from Eq.\,(\ref{pressurelaw}). On comparison with the value of $n$ valid for planar solid targets, we note that the difference could well reflect the change in target geometry from plane to spherical surfaces. The proof of the validity of this scaling law (Eq.\,(\ref{pressurelaw})) for the current geometry with respect to $\lambda$ and $\tau_p$ is left for future work.  \\

At lower values of energy-on-droplet $E_{\textrm{OD}}$ in Fig.\,\ref{fig:powerlaw}, below $\sim$0.2\,mJ, the data no longer follow the power law. This is due to the physics governing the \emph{onset} of the plasma formation upon the laser ablation of the metal.
The laser fluence at the threshold of ablation is given by \cite{Chichkov1996,Gamaly2002a} 
\begin{equation}
F_\textrm{th} = \rho \Delta H \sqrt{\kappa \tau_p} \approx 0.6\,\textrm{J\,cm}^{-2},
\label{threshold}
\end{equation}
with latent heat of vaporization $\Delta H \approx 2.2$\,MJ\,kg$^{-1}$ (taking the average of In and Sn in the mixture, both values being within 10\% of this average), and thermal diffusivity $\kappa \approx 16.4$\,mm$^2$/s \cite{Stankus2012}. At the onset, inverse bremsstrahlung does not yet play a role and we have to take into account the reflectivity $\mathcal{R}$ of the surface \footnote{The optical constants are, for pure tin in absence of laser light, $n=3.92$, $k=8.65$ (from \cite{Cisneros1982} at $1.2$\,eV at 523\,K) yielding a reflectivity of $\mathcal{R}=84$\%. For indium a reflectivity of $\mathcal{R}=91$\% is obtained from $n=1.84$, $k=8.38$ (from \cite{Golovashkin1967} at $1.1\,\mu$m at 295\,K). An estimate of the eutectic optical response is obtained from the averaging the reflectivities, yielding an absorption $1-\mathcal{R}=13$\%.}, and multiply Eq.\,(\ref{threshold}) with the factor $1/(1-\mathcal{R})$ \cite{gamaly2011physics}. For simplicity we take $\mathcal{R}$ to be constant during the duration of the pulse. From these considerations, we obtain the prediction $F_\textrm{th} \approx 5\,$J\,cm$^{-2}$, identical to the $\sim$5\,J\,cm$^{-2}$ from plasma modeling \cite{Masnavi2011}.
This threshold laser fluence translates into a minimum necessary pulse energy. 
Such a threshold energy can be straightforwardly included in our model by incorporating an offset pulse energy $E_{\textrm{OD,0}}$, such that the expression for the droplet velocity reads
\begin{equation}
U=\widetilde{B} \cdot (E_{\textrm{OD}}-E_{\textrm{OD,0}})^{\alpha}. \label{eq:powerlaw2}
\end{equation}
Here, the power $\alpha$ is taken as a free parameter since generally $v=v(I)$ \cite{Masnavi2011} and its scaling with intensity could influence the momentum scaling relation given by Eq.\,(\ref{eq:powerlaw1}). 
A single fit of Eq.\,(\ref{eq:powerlaw2}) to the full energy range of the concatenated data set (see Fig.\,\ref{fig:powerlaw}(a)) yields a power $\alpha=0.59(3)$, consistent with the postulated power of $5/9$. A proportionality constant $\widetilde{B}=35(5)$\,m\,s$^{-1}$\,mJ$^{-\alpha}$ is obtained similar to the result from the fit of Eq.\,(\ref{eq:powerlaw1}) above. For the offset we find $E_{\textrm{OD,0}}=0.05(1)$\,mJ, from which we in turn obtain $F_\textrm{th}=2.4(8)$\,J\,cm$^{-2}$ by dividing the offset energy by $\pi R_0^2$. This value is in reasonable agreement with the simple estimate that yielded $\sim$5\,J\,cm$^{-2}$. We conclude that all data are excellently described by a single fit of our model to the data, predicting the power-law scaling of $U$ with $E_{\textrm{OD}}$ in Eq.\,(\ref{eq:powerlaw2}) now including a threshold energy.

\subsubsection{Propulsion efficiency}
Having found an adequate description of the plasma propulsion mechanism over the complete measurement range, we can now derive an optimum condition for the laser-induced droplet propulsion. The momentum coupling coefficient $C_{\textrm{m}} \equiv p/I = (MU)/E$ \cite{Phipps1988} given by Eq.\,(\ref{pressurelaw}) is a figure of merit in plasma propulsion providing a measure for the propulsion efficiency in terms of total imparted momentum $MU$ per unit laser energy $E$. Efficient propulsion implies an efficient momentum to kick the droplet. This kick initiates the expansion of the droplet into a thin sheet (see below). Propulsion and expansion speeds are coupled, and their ratio is a function of the focusing conditions \cite{Gelderblom:2015}. Industrial needs could dictate the minimum size of the focus, as it influences the laser-to-droplet alignment stability. This will further influence the energy efficiency due to the geometrical overlap of laser focus and droplet. For both focusing conditions, however, we can define optimum conditions under which a minimum amount of laser pulse energy (on droplet) is required to reach a given velocity. Less energy is then available to produce, \emph{e.g.}, fast ionic and neutral debris that could limit machine lifetime \cite{Banine2011}. Given the offset power law (Eq.\,(\ref{eq:powerlaw2})), $C_\textrm{m}$ can be obtained as a function of pulse energy-on-droplet $E_{\textrm{OD}}$ (see Fig.\,\ref{fig:powerlaw}(b)) reaching a maximum at $E_{\textrm{OD,max}}=E_{\textrm{OD,0}}/(1-\alpha)$ at which point a minimum amount of energy is used to achieve a given velocity. A sequence of optimal pulses, spaced just a few $\tau_a$ apart to allow the plasma to recombine, could then be used to achieve a specified velocity. Of course, the relation between the energy-on-droplet $E_{\textrm{OD}}$ and total laser pulse energy $E$ is given by the geometrical overlap of laser and droplet. Our data (Fig.\,\ref{fig:powerlaw}(b)) thus indicate that the total energy efficiency is highest, at $E_{\textrm{OD,max}}$, for the smallest spot size used in this work.

\subsubsection{Indium-tin-plasma vs water-vapor propulsion}
Our description of the plasma propulsion of In-Sn micro-droplets shows striking analogies with recent work on water droplets \cite{Klein2015}. In that work, mm-sized water droplets are dyed to efficiently absorb 532\,nm light from a Nd:YAG laser pulse, at intensities well below the threshold for optical breakdown to avoid plasma generation. Instead, water vapor is expelled at its thermal speed and accelerates the droplet. In the dyed-water experiments, laser light is absorbed in a thin layer of thickness $\delta$, given by the optical penetration depth. This layer first needs to be heated from room temperature $T_0$ to boiling point $T_b$, yielding an expression for the threshold fluence $F_\textrm{th,w}=\rho_\textrm{w} \, \delta \, c \, (T_b-T_0)$ \cite{Klein2015}, where $c$ is the specific heat capacity of water and $\rho_\textrm{w}$ is its density. Any additional energy leads to vaporization of the liquid, expelling mass at a thermal velocity $u\approx 400\,$m/s which leads to the propulsion of the water droplet. When comparing the onset for propulsion of water droplets to the threshold fluence for In-Sn (Eq.\,(\ref{threshold})), one observes that for water, the mass affected by the laser is set by the optical penetration depth, whereas for In-Sn it is set by the thermal penetration depth $\sqrt{\kappa \tau_p}$. The energy density required for droplet propulsion is given by the sensible heat for water, and by the latent heat $\Delta H$ for In-Sn. The In-Sn droplet surface is shielded from the laser light by the plasma that develops at time scales well shorter than the pulse length. Laser light is absorbed by this plasma layer, heating it to high temperatures (several 10\,eV). As a result, a relatively small amount of mass ablated is expelled at high velocities $v \gg u$ thus mass-efficiently propelling the droplet. In conclusion, even though the physical mechanism behind the propulsion of In-Sn and water droplets is different, both processes are described by the forceful directive expulsion of material at a given velocity from a thin layer facing the laser light.\\

As we will demonstrate below, the fluid dynamic response to the laser impact described here and in Ref.\,\cite{Klein2015} is completely independent from the exact mechanism of propulsion. Mapping the system to the dimensionless numbers governing fluid dynamics allows for an identical treatment of said response for indium-tin and water.

\subsection{Droplet deformation}
\begin{figure}[t]
	\includegraphics[width=0.48\textwidth]{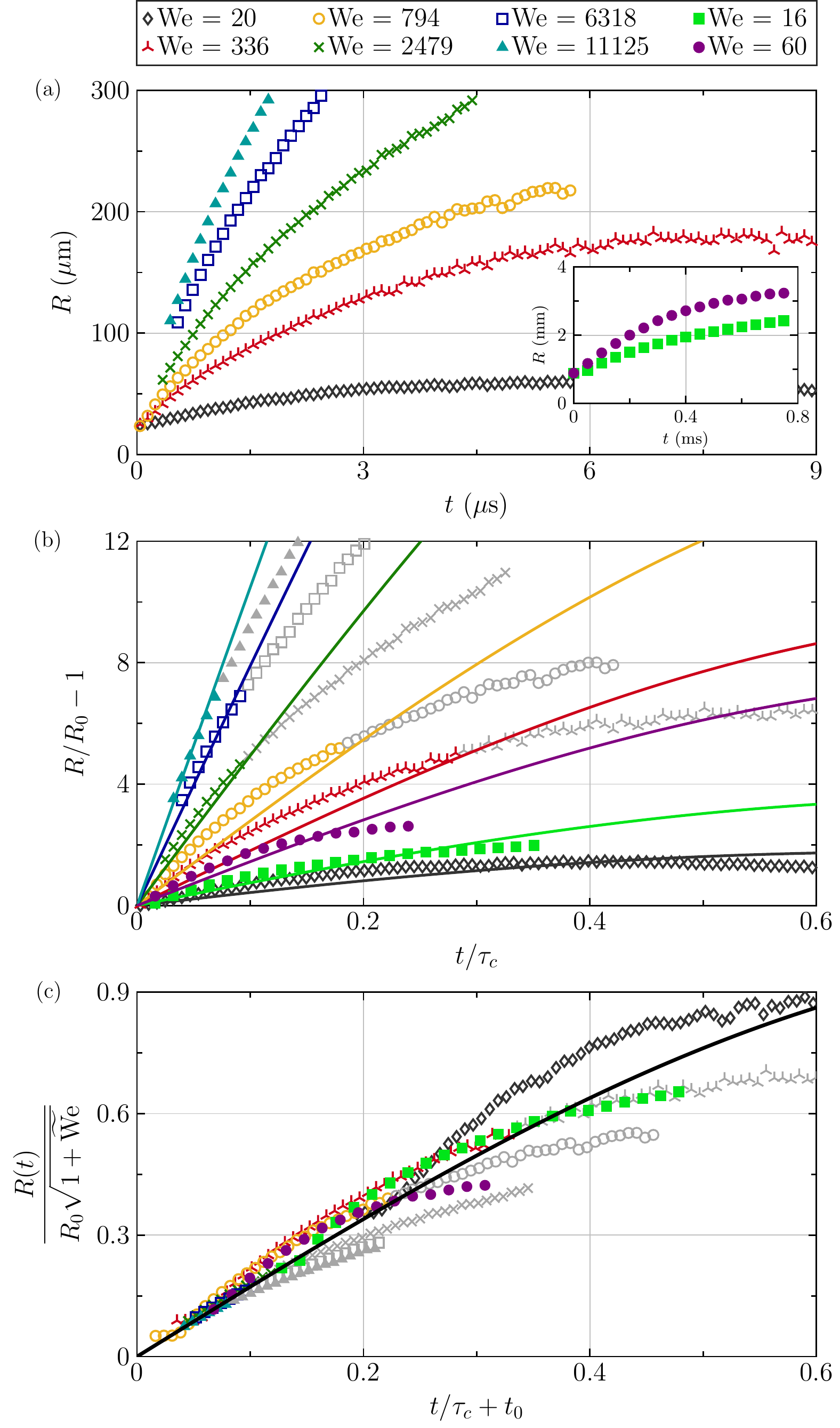}
	\caption{Radial expansion of In-Sn and water droplets as function of time. (a) Dimensional plot of the experimental data for In-Sn droplets determined from shadowgraphy (see Fig.\,\ref{fig:droplets}), with Weber numbers calculated using the droplet velocities from Fig.\,\ref{fig:powerlaw}. The inset shows data for mm-sized water droplets taken from a previous study \cite{Klein2015}. (b) Experiments (markers) and theory (solid lines, Eq.\,(\ref{eq:theory})) in a dimensionless representation. The data points are depicted in gray from the time onward at which formation of ligaments becomes apparent and comparison with the present theory is no longer relevant. Comparison between model and water data taken from \cite{Gelderblom:2015}. (c) Experimental data (markers) and theory (solid line) collapsed to a universal curve using a scaled Weber number $\widetilde{\textrm{We}}$ and time offset $t_0$ (see main text).
	\label{fig:r_tc}}
\end{figure}%
The expansion of the droplet for the 115\,$\mu$m focusing case is studied by changing the time delay between the pulse from the Nd:YAG laser and the shadowgraphy lasers in steps of 100\,ns. A stroboscopic movie of the deformation dynamics is thus obtained (see Fig.\,\ref{fig:droplets}) with an adequate temporal resolution. We measure the expanded radius of the droplet over time as shown in Fig.\,\ref{fig:r_tc}(a) using the side view shadowgraphs, in which the droplet motion is captured within the depth of field of the imaging optics at all time delays. In addition, the front view shadowgraphs are used to make sure the droplet expands into an axisymmetric shape. For comparison, we also include data from mm-sized water droplets, the deformation of which takes place on a time and length scale several orders of magnitude larger than for the In-Sn data (see inset in Fig.\,\ref{fig:r_tc}(a)). However, by appropriately rescaling the data by the initial droplet radius~$R_0$ as the characteristic length scale and the capillary time~$\tau_c= \sqrt{\rho R_0^3/\gamma}$ as the characteristic time scale for the droplet expansion and subsequent retraction, we can represent all data in one graph: Fig.\,\ref{fig:r_tc}(b). For indium-tin, $\tau_c\approx 14\, \mu$s given its surface tension \mbox{$\gamma=538$\,mN/m \cite{Pstrus2013}}. The expansion dynamics of the droplet can be described in terms of the Weber number $\textrm{We}=\rho R_0 U^2/\gamma$ \cite{Klein2015,Gelderblom:2015}, a dimensionless number that is a measure for the relative importance of the droplet's kinetic energy compared to its initial surface energy.

In the limit where the droplet expands into a flat, thin sheet that subsequently recedes due to surface tension an analytical expression can be found for the radius~$R$ as function of time~$t$ (for the derivation \mbox{see \cite{Gelderblom:2015}})
\begin{eqnarray}
\frac{R(t)}{R_0}&=&\cos{(\sqrt{3} t/\tau_c)} + \nonumber \\  && \bigg( \frac{2}{3} \bigg)^{1/2} \bigg(\frac{E_{k,d}}{E_{k,cm}} \bigg)^{1/2} \textrm{We}^{1/2} \sin{(\sqrt{3} t/\tau_c)},
\label{eq:theory}
\end{eqnarray}
where the parameter $E_{k,d}/E_{k,cm}$ is the partition of the total kinetic energy given to the droplet by the laser impact into a deformation kinetic energy $E_{k,d}$ and propulsion kinetic energy of the droplet's center-of-mass $E_{k,cm}$.
This parameter depends on the laser beam profile and can be determined analytically \cite{Gelderblom:2015}.
For all practical purposes in the present study, the laser beam profile can be considered flat, which gives $E_{k,d}/E_{k,cm}=0.5$\,\cite{Gelderblom:2015}.
The water data \cite{Klein2015} was obtained with a relatively more focused laser beam, for which $E_{k,d}/E_{k,cm}=1.8$ was calculated.
Figure \,\ref{fig:r_tc}(b) shows that the model prediction is in good agreement with the experimental data taking into account that it is derived from first principles and does not incorporate any fitting parameter.
In the comparison we truncated the experimental data where the formation of ligaments becomes apparent in the front view images, as at this point in time the assumption of a thin sheet of constant mass is clearly violated.\\

With the analytical expression at hand, the indium-tin and water data can be collapsed onto a universal curve by employing a trigonometric identity to rewrite Eq.\,(\ref{eq:theory}) to
\begin{equation}
\frac{R(t)}{R_0 \sqrt{1+\widetilde{\textrm{We}}}} = \sin{\left(\sqrt{3} (t/\tau_c + t_0)\right)} , \vspace{0.1mm} \label{eq:master}
\end{equation}
introducing an offset time $t_0$ and a scaled Weber number $\widetilde{\textrm{We}}$ that depends on the energy partition ratio $E_{k,d}/E_{k,cm}$,
\begin{eqnarray}
  \widetilde{\textrm{We}} &=& \frac{2}{3} \frac{E_{k,d}}{E_{k,cm}} \textrm{We} , \nonumber \\
  t_0 &=& \frac{1}{\sqrt{3}}\tan^{-1}\left(\widetilde{\textrm{We}}^{-1/2}\right). \nonumber
\label{eq:masterparams}
\end{eqnarray}
The data scaled with Eq.\,(\ref{eq:master}) indeed overlap completely and agree with the now single theoretical prediction (see Fig.\,\ref{fig:r_tc}(c)). This collapse demonstrates that the expansion dynamics, given the above scaling relation, are successfully described by a single dimensionless parameter. \\
 
\section{Conclusions}
We demonstrated that a micron-sized free-falling In-Sn micro-droplet hit by a high-intensity 10\,ns-long laser pulse is propelled by a plasma pressure ``kick''. The propulsion dynamics can be well understood in terms of a power law, that describes the dependence on the laser-pulse energy and incorporates an offset parameter to account for the threshold of plasma generation. 
This scaling law provides a useful tool to optimize the momentum coupling of the laser to the target. The description of the propulsion as a short recoil-pressure is similar to that presented in \cite{Klein2015}, where a millimeter-sized water droplet is accelerated by the directional emission of vapor upon the absorption of light from a short laser pulse. We find one-to-one correspondences between the two propulsion mechanisms, including the description of an analogous onset effect, even though the physical origins of the propulsion are different. Continuing its flight, the droplet expands until fragmentation occurs. By a proper rescaling we show that all indium-tin and water data can be collapsed onto a universal curve that is accurately described by a previously developed analytical model. The results and scaling laws presented in this work provide a tool to optimize and control the droplet shaping by laser pulse impact for EUV lithography applications.

\section*{Acknowledgments}
We thank the mechanical workshop, and the design, electronic, and software departments of AMOLF for the construction of the droplet generator setup. Further, we thank J. Broeils for contributing to the droplet analyses in the early stages of the experiment. T. Cohen Stuart is thanked for technical assistance. H.G. and A.L.K. acknowledge financial support by FOM.

\bibliography{bib}

\end{document}